\documentclass[prl,showpacs,twocolumn,superscriptaddress,longbibliography]{revtex4}

\usepackage{amssymb,amsmath}
\usepackage{graphicx}% Include figure files
\usepackage{dcolumn}% Align table columns on decimal point
\usepackage{bm}% bold math
\usepackage{epstopdf}

\newcommand{\eps}{\varepsilon}

\newcommand{\rmd}{\mathrm{d}}

\begin{document}

\title{Strong Purcell effect in anisotropic $\varepsilon$-near-zero metamaterials}

\author{A.V. Chebykin}
\author{A.A. Orlov}
    \affiliation{ ITMO University, 49 Kronverkskiy pr., St. Petersburg 197101, Russia}
\author{A.S. Shalin}
    \affiliation{ ITMO University, 49 Kronverkskiy pr., St. Petersburg 197101, Russia}
\affiliation{Ulyanovsk Branch of Kotel'nikov Institute of Radio Engineering and Electronics, Goncharov Str. 48,
432011 Ulyanovsk, Russia}
\affiliation{Ulyanovsk State University, L. Tolstoy str. 42, Ulyanovsk, Russia}
\author{A.N. Poddubny}\email{poddubny@coherent.ioffe.ru}
    \affiliation{ ITMO University, 49 Kronverkskiy pr., St. Petersburg 197101, Russia}
    \affiliation{Ioffe Institute of the RAS, 26 Politekhnicheskaya st., St. Petersburg 194021, Russia}
\author{P.A. Belov}
    \affiliation{ ITMO University, 49 Kronverkskiy pr., St. Petersburg 197101, Russia}

 \begin{abstract}
We theoretically demonstrate the strong Purcell effect in $\varepsilon$-near-zero ultra-anisotropic uniaxial metamaterials with elliptic isofrequency surface. Contrary to the hyperbolic metamaterials, the effect does not rely on the diverging density of states and evanescent waves. As a result, both the radiative decay rate and the far-field emission power are enhanced. The effect can be realized in the periodic layered metal-dielectric nanostructures with complex unit cell containing two different metallic layers.
\end{abstract}


\maketitle

Epsilon-near-zero materials, i.e. materials with small effective permittivity $\varepsilon$, are in the focus of active theoretical~\cite{basharin2013,PhysRevB.89.085105} and experimental~\cite{liu2008,Vesseur2013,maas2013} research being very promising for applications including, for example,  guiding of light~\cite{edwards2008,luo2012}, phase front manipulation~\cite{silveirinha2007}, optical circuitry~\cite{metactronics}, and photovoltaics~\cite{molesky2013}. However, at the first glance they do not look beneficial for  nanophotonic applications and enhancement of  the light-matter coupling. Indeed, the Purcell factor in an isotropic medium with the dielectric constant $\eps$ is equal to
$\sqrt{\eps}$
and tends to zero for small $\eps$~\cite{Novotny2006}. In this letter, we draw attention to the {\it anisotropic} uniaxial $\eps$-near-zero medium,
satisfying the condition
\begin{equation}\label{eq:cond}
0<\eps_{xx}=\eps_{yy}\ll\eps_{zz}\:.
\end{equation}
The Purcell factor for the emitter embedded in the uniaxial material and polarized perpendicular to the symmetry axis $z$ is equal to \cite{poddubny2011pra}
\begin{equation}
F_{{\rm purc},x}=F_{{\rm purc},y}=\frac{\eps_{zz}}{\sqrt{\eps_{xx}}}+\frac{3\sqrt{\eps_{xx}}}{4}\:.\label{eq:1}
\end{equation}
In the isotropic regime $\eps_{xx}=\eps_{zz}\equiv\eps$ Eq.~\eqref{eq:1} reduces to $\sqrt{\eps}$. However, in the strongly anisotropic limit \eqref{eq:cond} one can obtain very large values of the Purcell factor  due to the divergence in the first term $\propto \eps_{zz}/\sqrt{\eps_{xx}}$. The condition Eq.~\eqref{eq:cond} can be realized when the longitudinal dielectric tensor component $\eps_{zz}$ stays finite and the transverse one ($\eps_{xx}$) becomes small. Alternatively, one can consider a situation when both $\eps_{zz}$ and $\eps_{xx}$ tend to zero with different rates, so that $\eps_{zz}$ remains much larger than $\eps_{xx}$. 

\begin{figure}[b]
\centering{\includegraphics[width=0.5\textwidth]{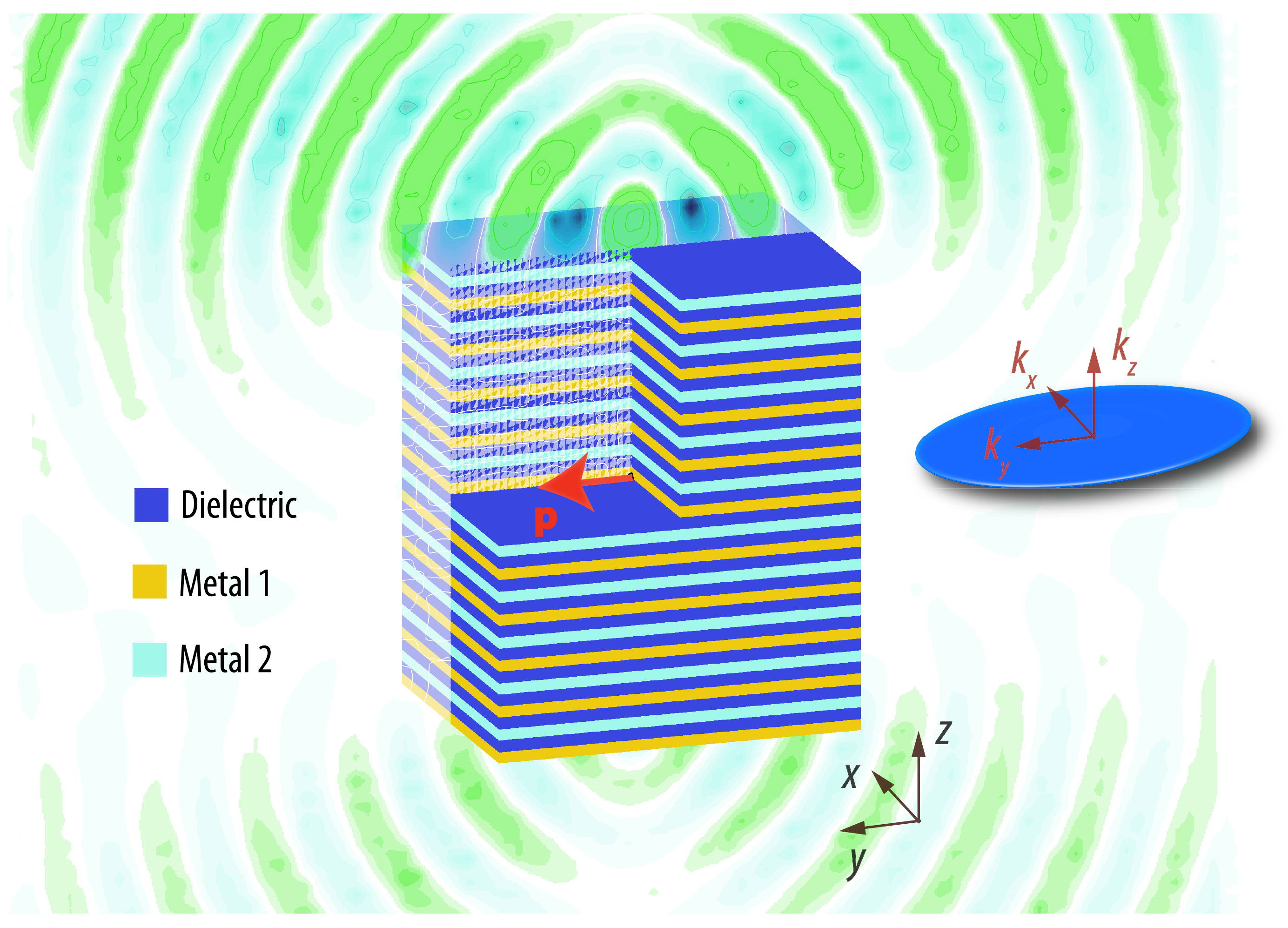}}
\caption{(Color online) Ultra-anisotropic elliptic metamaterial realized as a plasmonic multilayer structure with four layers per unit cell.
The color map illustrates the distribution of the $z$ component of the electric field induced by the dipole oriented along the layers and placed inside the structure (yellow arrow). Calculation has been performed at the frequency $538$~THz that corresponds to the regime with  ultra-oblate elliptic isofrequency surface illustrated in the inset.
\label{fig:1}
}
\end{figure}

In the  media satisfying Eq.~\eqref{eq:cond} the isofrequency surface, i.e. the surface spanned by the wave vectors $k$ of transverse magnetic (TM) modes corresponding to the same frequency, has a shape of a strongly oblate ellipsoid (see the inset of Fig.~\ref{fig:1}). The condition~\eqref{eq:cond} is realized at the elliptic side of the  topological transition between elliptic and hyperbolic regimes~\cite{shalaev2011,cortes2012,Drachev2013,poddubny2013NatPhot}. While the hyperbolic regime, when $\eps_{xx}<0$, allows to realize strong Purcell effect as well, it is strongly qualitatively different from the considered case. Particularly, in the hyperbolic metamaterials the spontaneous emission  is mostly due to the photon modes with large wave vectors, that lead to the diverging density of states but are evanescent outside the structure and can not be detected in the far field unless the structure surface is patterned to facilitate their outcoupling~\cite{lu2014}. Hence, the transition from elliptic to the hyperbolic regime is accompanied by shortening of the emission lifetime  and {\it suppression} of the observed far field emission intensity~\cite{tumkur2011,Kim2012}.  In the considered anisotropic elliptic case
the density of states stays finite, no evanescent waves are involved, and, as a result, both the radiative decay rate and the far field emission power can be enhanced.

%%%%%%%%%%%%%%%%%%%%%%%%%%%%%%%%%%%%%%%%

%%%%%%%%%%%%%%%%%%%%%%%%%%%%%%%%%%%%%%%%

Now we proceed to the discussion of the origin of the strong Purcell effect and the realization of the anisotropic elliptic regime  Eq.~\eqref{eq:cond} in  layered metal-dielectric metamaterials.
The origin of the spontaneous emission enhancement  can be most easily demonstrated by the Fermi Golden rule calculation:
\begin{equation}
\frac{1}{\tau}=\frac{2\pi}{\hbar}\sum\limits_{\bm k,\sigma}|\bm d\cdot\bm E_{\bm k,\sigma}|^{2}\delta(\hbar\omega_{\bm k,\sigma}-\hbar \omega_{0})\:.\label{eq:Fermi}
\end{equation}
Here, $\tau$ is the radiative decay rate, $\omega_{0}$ is the emission frequency for the two-level system, $\bm d$ is the dipole matrix element, and $\bm E_{\bm k,\sigma}$ is the electric field operator amplitude corresponding to the single quantum of radiation. The integration is performed over  the wave vectors $\bm k$ of  emitted waves with the TE or TM polarization denoted by $\sigma$.  Below we outline  the calculation of the spontaneous emission rate; the general result  valid both in elliptic and hyperbolic regimes and accounting for losses can be found in Ref.~\cite{poddubny2011pra}.
The  electric field  amplitude can be presented as
\begin{equation}
\bm E_{\bm k,\rm TM}=  \sqrt{\frac{2\pi\hbar\omega_{\bm k,\rm TM}}{{V_{{\rm mode,TM},\bm k}}}}
\left(\cos\theta_{\bm k}\hat{\bm \varphi}-\frac{\eps_{xx}}{\eps_{zz}}\sin\theta_{\bm k}\hat{\bm z}\right)\:,
\end{equation}
where the effective mode volume $V_{\rm mode}$ is determined from the quantization condition for the plane waves $V\bm E_{\bm k}(\hat\eps\bm E_{\bm k})=2\pi\hbar \omega_{\bm k,\rm TM}$, $V$ is the normalization volume, $\theta$ and $\varphi$ are the spherical coordinates of the wave vector $\bm k$. The effective mode volume 
\begin{equation}
V_{{\rm mode,TM},\bm k}=V\eps_{xx}^{2}/n^{2}_{\bm k,{\rm TM}}(\theta_{\bm k})\label{eq:V}
\end{equation}
can be expressed via  effective refractive index $n_{\rm TM}$ determining the TM  modes dispersion $\omega_{k,\rm TM}=ck/n_{\rm TM}(\theta_{\bm k})$,
\begin{equation}
n_{\bm k,{\rm TM}}(\theta_{\bm k})=\left(\frac{\sin^{2}\theta_{\bm k}}{\eps_{zz}}+\frac{\cos^{2}\theta_{\bm k}}{\eps_{xx}}\right)^{-1/2}\:.\label{eq:n}
\end{equation}
Substituting Eqs.~\eqref{eq:V},\eqref{eq:n} into Eq.~\eqref{eq:Fermi} and performing the integration over $k$ and $\varphi$ we obtain  the  contribution of the TM modes to the radiative decay rate for the $x$-polarized transition\:
\begin{multline}
\frac{1}{\tau_{x,\rm TM}}=\frac{d^{2}}{2\hbar}\left(\frac{\omega_{0}}{c}\right)^{3}\int\limits_{0}^{\pi}\frac{\rmd\theta_{\bm k}\sin\theta_{\bm k}\cos^{2}\theta_{\bm k} n_{\rm TM}^{5}(\theta_{\bm k})}{\eps_{xx}^{2}}\\=
\frac{d^{2}}{3\hbar}\left(\frac{\omega_{0}}{c}\right)^{3}\frac{\eps_{zz}}{\sqrt{\eps_{xx}}}\:.\label{eq:theta1}
\end{multline}
Equation~\eqref{eq:theta1} has a divergency for $\eps_{xx}\to +0$. While this divergency takes place at the threshold of the hyperbolic regime the origin of the emission enhancement is quite different from the case of hyperbolic metamaterials, because the density of states stays finite. Particularly, in the ultra-anisotropic regime $(\eps_{xx}\ll \eps_{zz})$ the spontaneous emission is dominated by the waves propagating at the grazing angles to the symmetry plane, $\theta_{\bm k}\approx \pi/2$:
\begin{equation}
\frac{1}{\tau_{x,\rm TM}}=\frac{d^{2}}{2\hbar} \left(\frac{\omega_{0}}{c}\right)^{3}\sqrt{\eps_{xx}}\int\limits_{-\infty}^{\infty}  \frac{\psi^{2} \rmd\psi}{[\psi^{2}+\eps_{xx}/\eps_{zz}]^{5/2}}\:,\label{eq:theta2}
\end{equation}
where $\psi=\theta_{\bm k}-\pi/2$\:. The sharp strong maximum in Eq.~\eqref{eq:theta2} for $\psi\sim \sqrt{\eps_{xx}/\eps_{zz}}\ll1$ is due to the diverging effective refractive index Eq.~\eqref{eq:n} which means vanishing effective mode volume Eq.~\eqref{eq:V}. The small effective mode volume leads to the strong Purcell effect, similarly to the case of resonant cavities.

The final result for Purcell factor is obtained by including the contribution of the TE modes (determined by the effective index $n_{\rm TE}=\sqrt{\eps_{xx}}$) and normalizing to the free-space radiative decay rate $1/\tau=4d^{2}\omega^{3}/(3\hbar c^{3})$. The expressions
for  $x$ and $z$ polarized emitters read 
\begin{align}
F_{{\rm purc},x}=F_{{\rm purc},y}&=\frac{\eps_{zz}}{\sqrt{\eps_{xx}}}+\frac{3\sqrt{\eps_{xx}}}{4}\:,\label{eq:x}\\
%:
%:
F_{{\rm purc},z}&=\sqrt{\eps_{xx}}\:.\label{eq:z}
\end{align}
It is the first term in Eq.~\eqref{eq:x} that presents the contribution of the TM waves to the spontaneous emission rate and can be arbitrary large in the ultra-anisotropic elliptic regime. The effect is present only for the emitters polarized in $xy$ plane, cf. Eq.~\eqref{eq:x} and Eq.~\eqref{eq:z}.

\begin{figure}
\center{\includegraphics[width=0.48\textwidth]{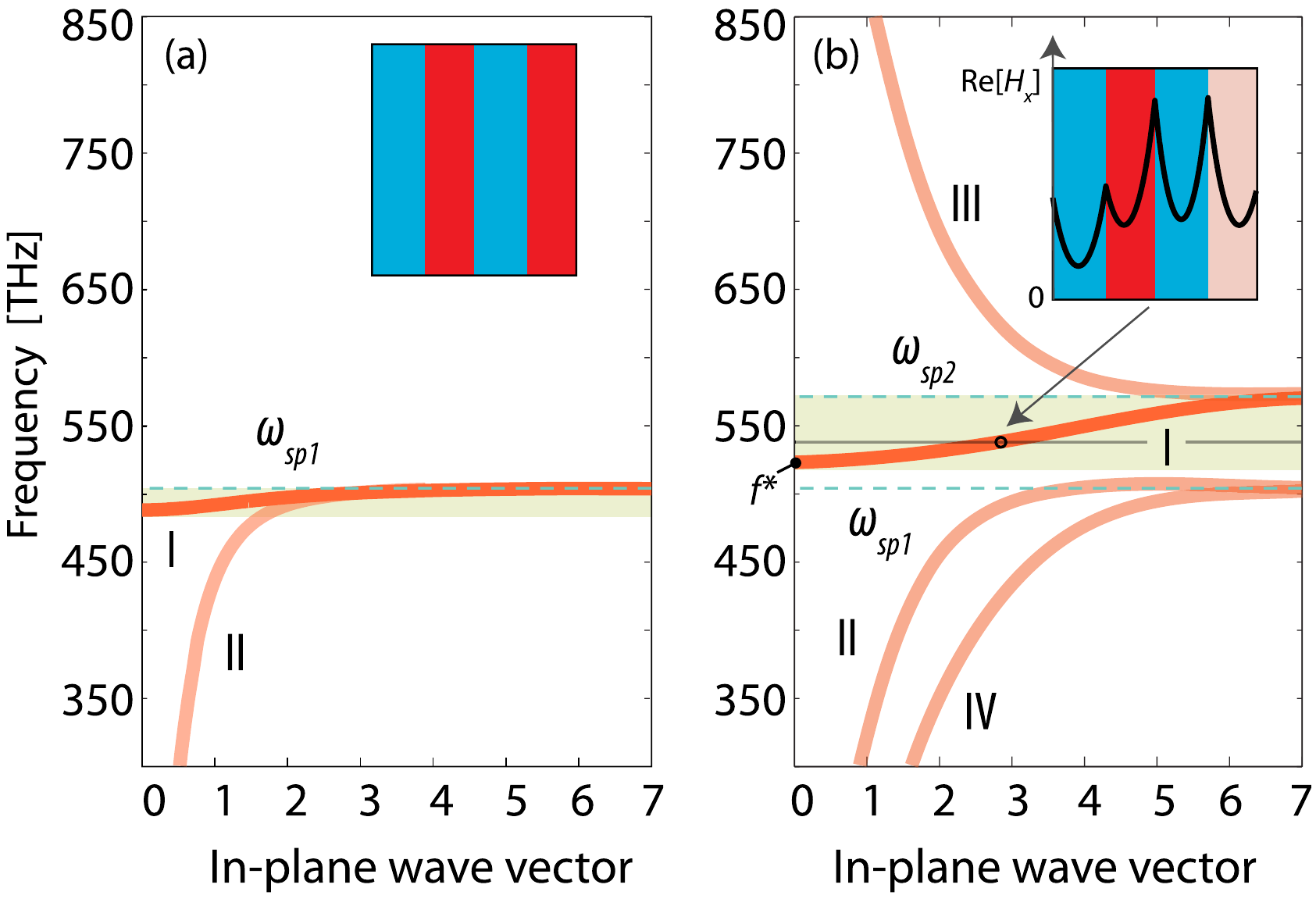}}
\caption{(Color online) Dispersion of the TM waves in  (a) conventional and 
(b) bi-periodic plasmonic multilayer at $k_{z}=0$. Horizontal green dashed lines show the surface plasmon frequencies $f_{sp1}$ and $f_{sp2}$ for  two types of metal-dielectric interfaces. The highlighted branch I corresponds to the flat elliptic isofrequency contour. Insets show the unit cells of the structures; the one of panel (b) contains the mode field profile at the frequency $f=538$~THz. Dielectric layers are shown by the blue color, metallic layers by red and pink colors. The in-plane wave vector has been normalized by multiplying over (a) $(d_{\rm me}+d_{\rm diel})/\pi$ and (b) $2(d_{\rm me}+d_{\rm diel})/\pi$. The calculation parameters are indicated in  text.
}\label{fig:2}
\end{figure}

This is how we arrive to an idea such the regime can be realized in actual artificial media. The seemingly natural approach is to consider plasmonic multilayers~\cite{subramania2012,orlovPNFA} that are layered metal-dielectric structures formed by a periodic stack of metallic and dielectric layers. Conventional multilayers formed by the layers of two kinds described typically with the following permittivities~\cite{AgranovichSuperlat}: 
\begin{align}
\eps_{xx}^{(\rm eff)}=
\eps_{yy}^{(\rm eff)}=\langle \varepsilon(z)\rangle\equiv
\frac{\eps_{\rm me}d_{\rm me}+\eps_{\rm diel}d_{\rm diel}}{d_{\rm me}+d_{\rm diel}},\\
\eps_{zz}^{(\rm eff)}=\langle \varepsilon^{-1}(z)\rangle^{-1}\equiv
\left(\frac{d_{\rm me}/\eps_{\rm me}+d_{\rm diel}/\eps_{\rm diel}}{d_{\rm me}+d_{\rm diel}}\right)^{-1}\:,
\end{align}
where the angular brackets denote the spatial averaging.
Analyzing these equations we find that the condition Eq.~\eqref{eq:cond} is realized at the frequency slightly above the transition frequency $\omega^{*}$, corresponding to the transition between the elliptic and hyperbolic regimes
\begin{equation}
\eps_{\rm me}(\omega^{*})=-\frac{d_{\rm diel}}{d_{\rm me}} \eps_{\rm diel}(\omega^{*})\:,\label{eq:elcond}
\end{equation}
when $\eps_{xx}^{(\rm eff)}$ turns to zero. The condition $\eps_{zz}^{(\rm eff)}(\omega^{*}) \ge 0$ requires $d_{\rm me}\le d_{\rm diel}$, i.e. the metallic layers should be thinner than the dielectric ones. Particularly, this means that the absolute value of $\eps_{\rm me}$ is larger than $\eps_{\rm diel}$ at the frequency $\omega^{*}$ and the frequency $\omega^{*}$ is lower  than the surface plasmon frequency determined from the condition
$\eps_{\rm me}=-\eps_{\rm diel}$.  However, it turns out that in this case the corresponding isofrequency contour consists not only of the elliptic contour but possesses an additional hyperbolic branch due to the inherent strong spatial dispersion of the plasmonic multilayer~\cite{orlovPRB}. 
Dispersion diagram of the structure as a function of the normalized in-plane wave vector $k_{y}$ is shown in Fig.~\ref{fig:2}(a). The mode I has an elliptic dispersion, but it is spectrally overlapped with the mode II that is hyperbolic one.
Hence, the effect of the mode I on the spontaneous emission is completely masked by the contribution of the mode II.

In order to obtain the isolated elliptic isofrequency contour we propose plasmonic multilayer structures having complex unit cell with bi-periodicity~\cite{orlovPRA}. Particularly, we consider the system formed by four layers per period, two different metallic ones and two equal dielectric ones, as shown in Fig.~\ref{fig:1}. We choose the dielectric layers with the same permittivity of $\varepsilon_d = 4.6$. The metallic layers are described by the Drude model: $\varepsilon_{me1,me2} = 1 - \lambda^2/\lambda_{p1,p2}^2$, with different plasma wavelengths $\lambda_{p1}$ = 250 nm and $\lambda_{p2}$ = 220 nm. All layers are chosen to have equal thickness of $31$~nm. Dispersion diagram  for this case is presented in Fig.~\ref{fig:2}(b). The 
photonic band structure of the bi-periodic multilayer  possesses  four branches, contrary to two branches in the conventional plasmonic multilayer with simple unit cell [Fig.~\ref{fig:2}(a)]. 

\begin{figure}
\center{\includegraphics[width=0.45\textwidth]{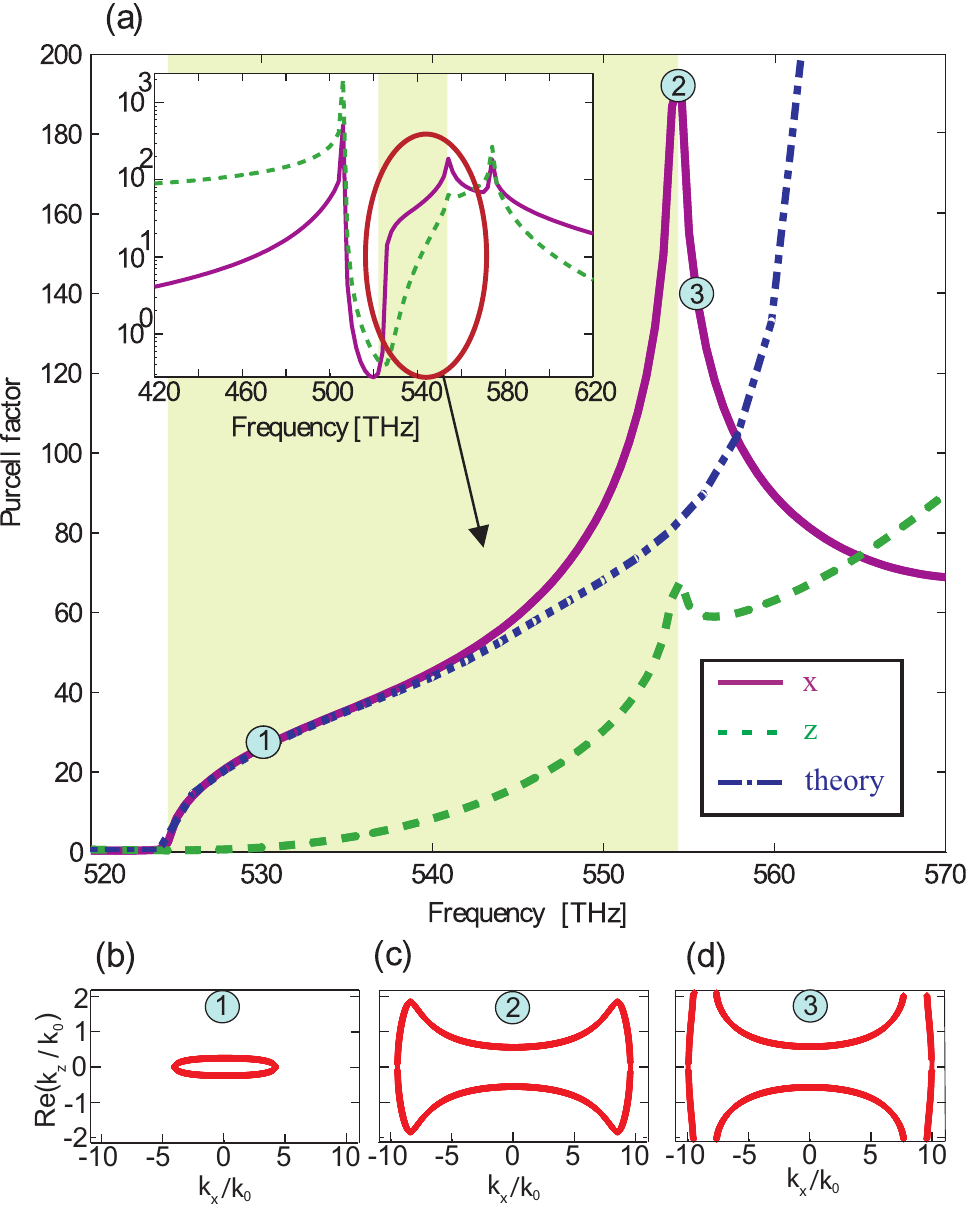}}
\caption{(Color online) (a) Dependence of the Purcell factor on the frequency. Existence domain of the mode I is highlighted. Inset shows the same curve in the logarithmic scale in a wider spectral range. Violet solid and green dashed curves are calculated for the  dipole oriented along the layers ($\parallel x$) and perpendicular to the layers ($\parallel z$),  respectively. Blue dash-dotted curve shows the effective medium result Eq.~\eqref{eq:x} corrected by the local field factor 1.4. Panels (b)--(d) present the isofrequency contours at  530, 554, 556 THz, respectively.
  }\label{fig:3}
\end{figure}

Since the complex unit cell is formed by  different metallic layers, two surface plasmon resonances can be distinguished in Fig.~\ref{fig:2}(b), positioned at the frequencies $f_{sp1}$ and $f_{sp2}$.  The mode I lies  between these two frequencies being bounded from above by $f_{sp2}$ and from below by the threshold frequency $f^*$ .
It is blue-shifted in comparison with the mode I of Fig.~\ref{fig:2}(a). At the same time, the mode II occupies the former region below $f_{sp1}$. Thus, the mode I becomes spectrally isolated allowing us to realize the ultra-anisotropic elliptic regime. Particularly, its magnetic field keeps the same sign within the unit cell [inset of Fig.~\ref{fig:2}(b)]. Hence, this mode is still qualitatively described within the effective medium approximation by the period-averaged effective dielectric constant $\varepsilon_{xx}=\langle \varepsilon(z)\rangle$ and remains elliptic [see Fig.~\ref{fig:3}(b)]. In the spectral range of mode I we expect strong radiative Purcell effect.

In order to demonstrate the effect, we have calculated the spectral dependence of the Purcell factor for the emitter placed in the middle of the dielectric layer inside the infinite periodic structure. We have applied the Green function technique for layered structures~\cite{tomas1999}. The result is shown in Fig.~\ref{fig:3}. The solid violet curve has been calculated for the  emitter oriented  along the layers, the green dashed curve corresponds to the perpendicular orientation. Figures~\ref{fig:3}(b)--(d) show the isofrequency contours  for three particular frequencies. The frequency  530 THz corresponds to the main result of this work: in this case  the  isofrequency contour has strongly anisotropic elliptic shape, and the  Purcell factor reaches  the value of 25 (violet line). For frequencies lower than 530 THz the most part of the isofrequency contour lies within the range of free space propagating waves, $k_{y}<k_{0}=\omega/c$, and hence the energy can be radiated in the far field [Fig.~\ref{fig:1}]. The increase of the Purcell factor in the elliptic regime is observed only for the emitter oriented parallel to the layers. Indeed, the elliptic regime is achieved in the frequency range 524 -- 540 THz, when the Purcell effect for the transverse emitter's orientation  is quite weak (dashed green curve). Such polarization dependence of the Purcell factor  is in perfect  qualitative agreement with the effective medium prediction Eqs.~\eqref{eq:x},\eqref{eq:z}. The  dash-dotted blue curve in panel (a) shows the effective medium result obtained using the values of $\eps_{xx}$ and $\eps_{zz}$  extracted by fitting the isofrequency contours. The function Eq.~\eqref{eq:x} has been multiplied by the  factor 1.4, that can be interpreted as a local field correction to the effective medium model~\cite{poddubny2012cross}. This semi-analytical expression 
 well describes the numerically calculated  frequency dependence of the Purcell factor [cf. violet and blue curves in Fig.~\ref{fig:3}(a) ]. 

Here, we have focused on the case of vanishing losses. While the Purcell enhancement due to the elliptic branch I is strongly suppressed for realistic losses, the effect can be further optimized by considering the metals with larger difference of the plasma frequencies. In the spectral range 540--554 THz the isofrequency contour evolves from an ellipsoid to the dumbbell, and, at frequencies larger than 554 THz it splits into two hyperbolic branches corresponding to mode III in Fig.~\ref{fig:2}. In these regimes the spontaneous emission is quite fast, but dominated by evanescent waves. 

To summarize, we have demonstrated how the strong Purcell enhancement of both the spontaneous emission rate  and the far-field emission power can be realized in the ultra-anisotropic uniaxial metamaterials where the transverse component of the dielectric tensor is positive but much smaller than the axial one. Our work shows that the  possibilities to engineer the photon dispersion and the light-matter coupling in layered metal-dielectric nanostructures reach far beyond the established concepts of  individual surface plasmon modes,  hyperbolic metamaterials or isotropic $\eps$-near-zero medium and are yet to be fully explored.

\acknowledgements
The authors are grateful to I.V. Shadrivov  and Yu. S. Kivshar for useful discussions. 
This work has been supported by the President of Russian Federation (Grant SP-2154.2012.1), the Government of Russian Federation (Grant 074-U01), and Russian Foundation for Basic Research (Project 14-02-31720). 
ANP acknowledges the support of the ``Dynasty'' foundation. The work of ASS 
(numerical simulations and investigating of the field distributions) has been funded by the Russian Science Foundation Grant No. 14-12-01227.

%\bibliography{blin}

\begin{thebibliography}{26}
\expandafter\ifx\csname natexlab\endcsname\relax\def\natexlab#1{#1}\fi
\expandafter\ifx\csname bibnamefont\endcsname\relax
  \def\bibnamefont#1{#1}\fi
\expandafter\ifx\csname bibfnamefont\endcsname\relax
  \def\bibfnamefont#1{#1}\fi
\expandafter\ifx\csname citenamefont\endcsname\relax
  \def\citenamefont#1{#1}\fi
\expandafter\ifx\csname url\endcsname\relax
  \def\url#1{\texttt{#1}}\fi
\expandafter\ifx\csname urlprefix\endcsname\relax\def\urlprefix{URL }\fi
\providecommand{\bibinfo}[2]{#2}
\providecommand{\eprint}[2][]{\url{#2}}

\bibitem[{\citenamefont{Basharin et~al.}(2013)\citenamefont{Basharin, Mavidis,
  Kafesaki, Economou, and Soukoulis}}]{basharin2013}
\bibinfo{author}{\bibfnamefont{A.~A.} \bibnamefont{Basharin}},
  \bibinfo{author}{\bibfnamefont{C.}~\bibnamefont{Mavidis}},
  \bibinfo{author}{\bibfnamefont{M.}~\bibnamefont{Kafesaki}},
  \bibinfo{author}{\bibfnamefont{E.~N.} \bibnamefont{Economou}},
  \bibnamefont{and} \bibinfo{author}{\bibfnamefont{C.~M.}
  \bibnamefont{Soukoulis}}, \bibinfo{journal}{Phys. Rev. B}
  \textbf{\bibinfo{volume}{87}}, \bibinfo{pages}{155130}
  (\bibinfo{year}{2013}).

\bibitem[{\citenamefont{Savoia et~al.}(2014)\citenamefont{Savoia, Castaldi,
  Galdi, Al\`u, and Engheta}}]{PhysRevB.89.085105}
\bibinfo{author}{\bibfnamefont{S.}~\bibnamefont{Savoia}},
  \bibinfo{author}{\bibfnamefont{G.}~\bibnamefont{Castaldi}},
  \bibinfo{author}{\bibfnamefont{V.}~\bibnamefont{Galdi}},
  \bibinfo{author}{\bibfnamefont{A.}~\bibnamefont{Al\`u}}, \bibnamefont{and}
  \bibinfo{author}{\bibfnamefont{N.}~\bibnamefont{Engheta}},
  \bibinfo{journal}{Phys. Rev. B} \textbf{\bibinfo{volume}{89}},
  \bibinfo{pages}{085105} (\bibinfo{year}{2014}).

\bibitem[{\citenamefont{Liu et~al.}(2008)\citenamefont{Liu, Cheng, Hand, Mock,
  Cui, Cummer, and Smith}}]{liu2008}
\bibinfo{author}{\bibfnamefont{R.}~\bibnamefont{Liu}},
  \bibinfo{author}{\bibfnamefont{Q.}~\bibnamefont{Cheng}},
  \bibinfo{author}{\bibfnamefont{T.}~\bibnamefont{Hand}},
  \bibinfo{author}{\bibfnamefont{J.~J.} \bibnamefont{Mock}},
  \bibinfo{author}{\bibfnamefont{T.~J.} \bibnamefont{Cui}},
  \bibinfo{author}{\bibfnamefont{S.~A.} \bibnamefont{Cummer}},
  \bibnamefont{and} \bibinfo{author}{\bibfnamefont{D.~R.} \bibnamefont{Smith}},
  \bibinfo{journal}{Phys. Rev. Lett.} \textbf{\bibinfo{volume}{100}},
  \bibinfo{pages}{023903} (\bibinfo{year}{2008}).

\bibitem[{\citenamefont{Vesseur et~al.}(2013)\citenamefont{Vesseur, Coenen,
  Caglayan, Engheta, and Polman}}]{Vesseur2013}
\bibinfo{author}{\bibfnamefont{E.~J.~R.} \bibnamefont{Vesseur}},
  \bibinfo{author}{\bibfnamefont{T.}~\bibnamefont{Coenen}},
  \bibinfo{author}{\bibfnamefont{H.}~\bibnamefont{Caglayan}},
  \bibinfo{author}{\bibfnamefont{N.}~\bibnamefont{Engheta}}, \bibnamefont{and}
  \bibinfo{author}{\bibfnamefont{A.}~\bibnamefont{Polman}},
  \bibinfo{journal}{Phys. Rev. Lett.} \textbf{\bibinfo{volume}{110}},
  \bibinfo{pages}{013902} (\bibinfo{year}{2013}).

\bibitem[{\citenamefont{Maas et~al.}(2013)\citenamefont{Maas, Parsons, Engheta,
  and Polman}}]{maas2013}
\bibinfo{author}{\bibfnamefont{R.}~\bibnamefont{Maas}},
  \bibinfo{author}{\bibfnamefont{J.}~\bibnamefont{Parsons}},
  \bibinfo{author}{\bibfnamefont{N.}~\bibnamefont{Engheta}}, \bibnamefont{and}
  \bibinfo{author}{\bibfnamefont{A.}~\bibnamefont{Polman}},
  \bibinfo{journal}{Nat. Phot.} \textbf{\bibinfo{volume}{7}},
  \bibinfo{pages}{907–912} (\bibinfo{year}{2013}).

\bibitem[{\citenamefont{Edwards et~al.}(2008)\citenamefont{Edwards, Alu, Young,
  Silveirinha, and Engheta}}]{edwards2008}
\bibinfo{author}{\bibfnamefont{B.}~\bibnamefont{Edwards}},
  \bibinfo{author}{\bibfnamefont{A.}~\bibnamefont{Alu}},
  \bibinfo{author}{\bibfnamefont{M.~E.} \bibnamefont{Young}},
  \bibinfo{author}{\bibfnamefont{M.}~\bibnamefont{Silveirinha}},
  \bibnamefont{and} \bibinfo{author}{\bibfnamefont{N.}~\bibnamefont{Engheta}},
  \bibinfo{journal}{Phys. Rev. Lett.} \textbf{\bibinfo{volume}{100}},
  \bibinfo{pages}{033903} (\bibinfo{year}{2008}).

\bibitem[{\citenamefont{Luo et~al.}(2012)\citenamefont{Luo, Xu, Chen, Hou, Gao,
  and Lai}}]{luo2012}
\bibinfo{author}{\bibfnamefont{J.}~\bibnamefont{Luo}},
  \bibinfo{author}{\bibfnamefont{P.}~\bibnamefont{Xu}},
  \bibinfo{author}{\bibfnamefont{H.}~\bibnamefont{Chen}},
  \bibinfo{author}{\bibfnamefont{B.}~\bibnamefont{Hou}},
  \bibinfo{author}{\bibfnamefont{L.}~\bibnamefont{Gao}}, \bibnamefont{and}
  \bibinfo{author}{\bibfnamefont{Y.}~\bibnamefont{Lai}},
  \bibinfo{journal}{Appl. Phys. Lett.} \textbf{\bibinfo{volume}{100}},
  \bibinfo{pages}{221903} (\bibinfo{year}{2012}).

\bibitem[{\citenamefont{Alù et~al.}(2007)\citenamefont{Alù, Silveirinha,
  Salandrino, and Engheta}}]{silveirinha2007}
\bibinfo{author}{\bibfnamefont{A.}~\bibnamefont{Alù}},
  \bibinfo{author}{\bibfnamefont{M.~G.} \bibnamefont{Silveirinha}},
  \bibinfo{author}{\bibfnamefont{A.}~\bibnamefont{Salandrino}},
  \bibnamefont{and} \bibinfo{author}{\bibfnamefont{N.}~\bibnamefont{Engheta}},
  \bibinfo{journal}{Phys. Rev. B} \textbf{\bibinfo{volume}{75}},
  \bibinfo{pages}{155410} (\bibinfo{year}{2007}).

\bibitem[{\citenamefont{Engheta}(2007)}]{metactronics}
\bibinfo{author}{\bibfnamefont{N.}~\bibnamefont{Engheta}},
  \bibinfo{journal}{Science} \textbf{\bibinfo{volume}{317}},
  \bibinfo{pages}{1698} (\bibinfo{year}{2007}).

\bibitem[{\citenamefont{Molesky et~al.}(2013)\citenamefont{Molesky, Dewalt, and
  Jacob}}]{molesky2013}
\bibinfo{author}{\bibfnamefont{S.}~\bibnamefont{Molesky}},
  \bibinfo{author}{\bibfnamefont{C.~J.} \bibnamefont{Dewalt}},
  \bibnamefont{and} \bibinfo{author}{\bibfnamefont{Z.}~\bibnamefont{Jacob}},
  \bibinfo{journal}{Opt. Expr.} \textbf{\bibinfo{volume}{21}},
  \bibinfo{pages}{A96} (\bibinfo{year}{2013}).

\bibitem[{\citenamefont{Novotny and Hecht}(2006)}]{Novotny2006}
\bibinfo{author}{\bibfnamefont{L.}~\bibnamefont{Novotny}} \bibnamefont{and}
  \bibinfo{author}{\bibfnamefont{B.}~\bibnamefont{Hecht}},
  \emph{\bibinfo{title}{{P}rinciples of {N}ano-{O}ptics}}
  (\bibinfo{publisher}{Cambridge University Press}, \bibinfo{address}{New
  York}, \bibinfo{year}{2006}).

\bibitem[{\citenamefont{Poddubny et~al.}(2011)\citenamefont{Poddubny, Belov,
  and Kivshar}}]{poddubny2011pra}
\bibinfo{author}{\bibfnamefont{A.~N.} \bibnamefont{Poddubny}},
  \bibinfo{author}{\bibfnamefont{P.~A.} \bibnamefont{Belov}}, \bibnamefont{and}
  \bibinfo{author}{\bibfnamefont{Y.~S.} \bibnamefont{Kivshar}},
  \bibinfo{journal}{Phys. Rev. A} \textbf{\bibinfo{volume}{84}},
  \bibinfo{pages}{023807} (\bibinfo{year}{2011}).

\bibitem[{\citenamefont{{Jacob} and {Shalaev}}(2011)}]{shalaev2011}
\bibinfo{author}{\bibfnamefont{Z.}~\bibnamefont{{Jacob}}} \bibnamefont{and}
  \bibinfo{author}{\bibfnamefont{V.~M.} \bibnamefont{{Shalaev}}},
  \bibinfo{journal}{Science} \textbf{\bibinfo{volume}{334}},
  \bibinfo{pages}{463} (\bibinfo{year}{2011}).

\bibitem[{\citenamefont{Cortes et~al.}(2012)\citenamefont{Cortes, Newman,
  Molesky, and Jacob}}]{cortes2012}
\bibinfo{author}{\bibfnamefont{C.~L.} \bibnamefont{Cortes}},
  \bibinfo{author}{\bibfnamefont{W.}~\bibnamefont{Newman}},
  \bibinfo{author}{\bibfnamefont{S.}~\bibnamefont{Molesky}}, \bibnamefont{and}
  \bibinfo{author}{\bibfnamefont{Z.}~\bibnamefont{Jacob}},
  \bibinfo{journal}{Journal of Optics} \textbf{\bibinfo{volume}{14}},
  \bibinfo{pages}{063001} (\bibinfo{year}{2012}).

\bibitem[{\citenamefont{Drachev et~al.}(2013)\citenamefont{Drachev, Podolskiy,
  and Kildishev}}]{Drachev2013}
\bibinfo{author}{\bibfnamefont{V.~P.} \bibnamefont{Drachev}},
  \bibinfo{author}{\bibfnamefont{V.~A.} \bibnamefont{Podolskiy}},
  \bibnamefont{and} \bibinfo{author}{\bibfnamefont{A.~V.}
  \bibnamefont{Kildishev}}, \bibinfo{journal}{Opt. Express}
  \textbf{\bibinfo{volume}{21}}, \bibinfo{pages}{15048} (\bibinfo{year}{2013}).

\bibitem[{\citenamefont{Poddubny et~al.}(2013)\citenamefont{Poddubny, Iorsh,
  Belov, and Kivshar}}]{poddubny2013NatPhot}
\bibinfo{author}{\bibfnamefont{A.}~\bibnamefont{Poddubny}},
  \bibinfo{author}{\bibfnamefont{I.}~\bibnamefont{Iorsh}},
  \bibinfo{author}{\bibfnamefont{P.}~\bibnamefont{Belov}}, \bibnamefont{and}
  \bibinfo{author}{\bibfnamefont{Y.}~\bibnamefont{Kivshar}},
  \bibinfo{journal}{Nature Photonics} \textbf{\bibinfo{volume}{7}},
  \bibinfo{pages}{958} (\bibinfo{year}{2013}).

\bibitem[{\citenamefont{Lu et~al.}(2014)\citenamefont{Lu, Kan, Fullerton, and
  Liu}}]{lu2014}
\bibinfo{author}{\bibfnamefont{D.}~\bibnamefont{Lu}},
  \bibinfo{author}{\bibfnamefont{J.~J.} \bibnamefont{Kan}},
  \bibinfo{author}{\bibfnamefont{E.~E.} \bibnamefont{Fullerton}},
  \bibnamefont{and} \bibinfo{author}{\bibfnamefont{Z.}~\bibnamefont{Liu}},
  \bibinfo{journal}{Nature Nanotechnology} \textbf{\bibinfo{volume}{9}},
  \bibinfo{pages}{48} (\bibinfo{year}{2014}).

\bibitem[{\citenamefont{Tumkur et~al.}(2011)\citenamefont{Tumkur, Zhu, Black,
  Barnakov, Bonner, and Noginov}}]{tumkur2011}
\bibinfo{author}{\bibfnamefont{T.}~\bibnamefont{Tumkur}},
  \bibinfo{author}{\bibfnamefont{G.}~\bibnamefont{Zhu}},
  \bibinfo{author}{\bibfnamefont{P.}~\bibnamefont{Black}},
  \bibinfo{author}{\bibfnamefont{Y.~A.} \bibnamefont{Barnakov}},
  \bibinfo{author}{\bibfnamefont{C.~E.} \bibnamefont{Bonner}},
  \bibnamefont{and} \bibinfo{author}{\bibfnamefont{M.~A.}
  \bibnamefont{Noginov}}, \bibinfo{journal}{Appl. Phys. Lett.}
  \textbf{\bibinfo{volume}{99}}, \bibinfo{eid}{151115} (\bibinfo{year}{2011}).

\bibitem[{\citenamefont{Kim et~al.}(2012)\citenamefont{Kim, Drachev, Jacob,
  Naik, Boltasseva, Narimanov, and Shalaev}}]{Kim2012}
\bibinfo{author}{\bibfnamefont{J.}~\bibnamefont{Kim}},
  \bibinfo{author}{\bibfnamefont{V.~P.} \bibnamefont{Drachev}},
  \bibinfo{author}{\bibfnamefont{Z.}~\bibnamefont{Jacob}},
  \bibinfo{author}{\bibfnamefont{G.~V.} \bibnamefont{Naik}},
  \bibinfo{author}{\bibfnamefont{A.}~\bibnamefont{Boltasseva}},
  \bibinfo{author}{\bibfnamefont{E.~E.} \bibnamefont{Narimanov}},
  \bibnamefont{and} \bibinfo{author}{\bibfnamefont{V.~M.}
  \bibnamefont{Shalaev}}, \bibinfo{journal}{Opt. Express}
  \textbf{\bibinfo{volume}{20}}, \bibinfo{pages}{8100} (\bibinfo{year}{2012}).

\bibitem[{\citenamefont{Subramania et~al.}(2012)\citenamefont{Subramania,
  Fischer, and Luk}}]{subramania2012}
\bibinfo{author}{\bibfnamefont{G.}~\bibnamefont{Subramania}},
  \bibinfo{author}{\bibfnamefont{A.~J.} \bibnamefont{Fischer}},
  \bibnamefont{and} \bibinfo{author}{\bibfnamefont{T.~S.} \bibnamefont{Luk}},
  \bibinfo{journal}{Appl. Phys. Lett.} \textbf{\bibinfo{volume}{101}},
  \bibinfo{pages}{241107} (\bibinfo{year}{2012}).

\bibitem[{\citenamefont{Orlov et~al.}(2014{\natexlab{a}})\citenamefont{Orlov,
  Iorsh, Zhukovsky, and Belov}}]{orlovPNFA}
\bibinfo{author}{\bibfnamefont{A.~A.} \bibnamefont{Orlov}},
  \bibinfo{author}{\bibfnamefont{I.~V.} \bibnamefont{Iorsh}},
  \bibinfo{author}{\bibfnamefont{S.~V.} \bibnamefont{Zhukovsky}},
  \bibnamefont{and} \bibinfo{author}{\bibfnamefont{P.~A.} \bibnamefont{Belov}},
  \bibinfo{journal}{Photonics and Nanostructures -- Fundamentals and
  Applications} \textbf{\bibinfo{volume}{12}}, \bibinfo{pages}{213}
  (\bibinfo{year}{2014}{\natexlab{a}}), ISSN \bibinfo{issn}{1569-4410}.

\bibitem[{\citenamefont{Agranovich and Kravtsov}(1985)}]{AgranovichSuperlat}
\bibinfo{author}{\bibfnamefont{V.}~\bibnamefont{Agranovich}} \bibnamefont{and}
  \bibinfo{author}{\bibfnamefont{V.}~\bibnamefont{Kravtsov}},
  \bibinfo{journal}{Solid State Communications} \textbf{\bibinfo{volume}{55}},
  \bibinfo{pages}{85 } (\bibinfo{year}{1985}).

\bibitem[{\citenamefont{Orlov et~al.}(2011)\citenamefont{Orlov, Voroshilov,
  Belov, and Kivshar}}]{orlovPRB}
\bibinfo{author}{\bibfnamefont{A.~A.} \bibnamefont{Orlov}},
  \bibinfo{author}{\bibfnamefont{P.~M.} \bibnamefont{Voroshilov}},
  \bibinfo{author}{\bibfnamefont{P.~A.} \bibnamefont{Belov}}, \bibnamefont{and}
  \bibinfo{author}{\bibfnamefont{Y.~S.} \bibnamefont{Kivshar}},
  \bibinfo{journal}{Phys. Rev. B} \textbf{\bibinfo{volume}{84}},
  \bibinfo{pages}{045424} (\bibinfo{year}{2011}).

\bibitem[{\citenamefont{Orlov et~al.}(2014{\natexlab{b}})\citenamefont{Orlov,
  Krylova, Zhukovsky, Babicheva, and Belov}}]{orlovPRA}
\bibinfo{author}{\bibfnamefont{A.~A.} \bibnamefont{Orlov}},
  \bibinfo{author}{\bibfnamefont{A.~K.} \bibnamefont{Krylova}},
  \bibinfo{author}{\bibfnamefont{S.~V.} \bibnamefont{Zhukovsky}},
  \bibinfo{author}{\bibfnamefont{V.~E.} \bibnamefont{Babicheva}},
  \bibnamefont{and} \bibinfo{author}{\bibfnamefont{P.~A.} \bibnamefont{Belov}},
  \bibinfo{journal}{Phys. Rev. A} \textbf{\bibinfo{volume}{90}},
  \bibinfo{pages}{013812} (\bibinfo{year}{2014}{\natexlab{b}}).

\bibitem[{\citenamefont{Toma\ifmmode~\check{s}\else \v{s}\fi{} and
  Lenac}(1999)}]{tomas1999}
\bibinfo{author}{\bibfnamefont{M.~S.} \bibnamefont{Toma\ifmmode~\check{s}\else
  \v{s}\fi{}}} \bibnamefont{and}
  \bibinfo{author}{\bibfnamefont{Z.}~\bibnamefont{Lenac}},
  \bibinfo{journal}{Phys. Rev. A} \textbf{\bibinfo{volume}{60}},
  \bibinfo{pages}{2431} (\bibinfo{year}{1999}).

\bibitem[{\citenamefont{Poddubny et~al.}(2012)\citenamefont{Poddubny, Belov,
  Ginzburg, Zayats, and Kivshar}}]{poddubny2012cross}
\bibinfo{author}{\bibfnamefont{A.~N.} \bibnamefont{Poddubny}},
  \bibinfo{author}{\bibfnamefont{P.~A.} \bibnamefont{Belov}},
  \bibinfo{author}{\bibfnamefont{P.}~\bibnamefont{Ginzburg}},
  \bibinfo{author}{\bibfnamefont{A.~V.} \bibnamefont{Zayats}},
  \bibnamefont{and} \bibinfo{author}{\bibfnamefont{Y.~S.}
  \bibnamefont{Kivshar}}, \bibinfo{journal}{Phys. Rev. B}
  \textbf{\bibinfo{volume}{86}}, \bibinfo{pages}{035148}
  (\bibinfo{year}{2012}).

\end{thebibliography}

\end{document}